\newcommand {\bea}{\begin{eqnarray}}
\newcommand {\eea}{\end{eqnarray}}
\newcommand {\be}{\begin{equation}}
\newcommand {\ee}{\end{equation}}

\documentstyle[12pt]{article}
\begin{document}
{\hbox to\hsize{October 1996 \hfill IASSNS-HEP 96/108}}\par
{\hbox to\hsize{hep-th/9610252 \hfill UPR-722-T}}\par
\begin{center}
{\LARGE \bf Resolution of   \\[0.1in]
Cosmological Singularities} \\[0.3in]
{\bf Finn Larsen}\footnote{e-mail: larsen@cvetic.hep.upenn.edu}\\[.05in]
{\it Department of Physics and Astronomy \\
University of Pennsylvania \\
Philadelphia, PA 19104} \\[.3in]
and \\ [.1in] 
{\bf Frank Wilczek}\footnote{e-mail: wilczek@sns.ias.edu} \\[.05in]
{\it School of Natural Sciences\\
Institute for Advanced Study\\
Princeton, NJ 08540}\\[.15in]
\end{center}

\begin{abstract}

We show that a class of 3+1 dimensional
Friedmann-Robertson-Walker cosmologies can be 
embedded within a variety of solutions of string theory.  In some
realizations the apparent singularities associated with the big bang or
big crunch are resolved at non-singular horizons of
higher-dimensional quasi-black hole solutions (with compactified {\it real\/}
time); in others plausibly they are resolved at D-brane 
bound states having no conventional space-time interpretation.

\end{abstract}

\newpage

In the course of their
historic treatment of the collapse of a cloud of pressureless dust in
general relativity, Oppenheimer and Snyder
introduced a striking construction: the joining of an external
Schwarzschild solution to the collapse phase of a closed
Friedmann-Robertson-Walker, homogeneous and isotropic model 
universe~\cite{snyder}.
The possibility to connect a static exterior geometry with a dynamic
interior occurs because the character -- space-like {\it versus\/}
time-like -- of the Schwarzschild
$r$ and $t$ variables changes upon passage through
the horizon.  Indeed in this way even the interior of a
vacuum Schwarzschild black hole can be interpreted as a cosmological
solution, although a highly anisotropic one.  To be specific, if we
define $v\equiv \sqrt {8m} (2m-r)^{1\over 2}$ then slightly inside the
horizon the line element takes the form 
\be
ds^2 ~=~ -dv^2 + {v^2\over 16m^2} dt^2 + (2m)^2 d\Omega_2^2~,
\ee
which appears to represent a primitive crunch singularity of the spatial
dimension
$t$ as $v\rightarrow
0^+$.  In reality of course there is no curvature singularity, but
only a singularity of the $v$ `time' coordinate that is resolved in
the complete embedding.

Here we will discuss some analogous phenomena in string theory and
simple effective field theories derived from it.  As we shall see, in
this framework several new features appear: the larger number of
available space dimensions make it possible to embed a homogeneous and
isotropic model 3+1 dimensional model universe in an interesting, non-singular 
fashion; the dilaton, moduli, and antisymmetric
tensor fields can provide interesting forms of `matter'; 
and -- perhaps most challenging -- the possibility arises that cosmological 
singularities are resolved in an intrinsically string-theoretic 
way that has no conventional space-time interpretation as it involves
D-branes and their open string dressing. 
 
\bigskip

It is convenient to begin the discussion by reference to the five
dimensional Schwarzschild metric
\be
ds^2 = -(1-{\mu\over r^2})dt^2 +(1-{\mu\over r^2})^{-1}dr^2 +r^2 d\Omega^2_3  
\ee
Inside the horizon it can be written after obvious renaming of 
variables as
\bea
ds^2&=& e^{-4\sigma} dy^2 -e^{4\sigma}d\tau^2 +\tau^2 d\Omega^2_3
\label{eqn:5dsolution} \\
e^{-4\sigma}&=& {\mu\over \tau^2}-1
\label{eqn:sigma}
\eea
The range of $\tau$ is $0<\tau^2<\mu$. To obtain an interesting 
embedded cosmology along the lines mentioned above, we compactify $y$.
(Such a construction was also employed in~\cite{behrndt,poppe}.) 
This induces an effective 4-dimensional dilaton 
$e^{-2\phi_4}=e^{-2\sigma}$ that can be interpreted as an 
independent matter field. In string frame the dimensionally reduced 
metric is eq.~\ref{eqn:5dsolution}, with the $dy^2$ term omitted. 
In terms of the Einstein frame metric $g_{E\mu\nu}=e^{-2\phi_4}g_{S\mu\nu}$
the four dimensional cosmology becomes
\be
ds_E^2= -e^{2\sigma}d\tau^2 + e^{-2\sigma}\tau^2 d\Omega^2_3 
\label{eqn:basicline}
\ee
with $e^{-2\sigma}$ as in eq.~\ref{eqn:sigma}. We regard the Einstein
frame line element as privileged, both because it governs the response
of conventional test bodies and because it is invariant under the
standard duality transformations.
Friedmann coordinates are introduced as
\be
ds^2_E = -dt^2 + R(t)^2 d\Omega^2_3 
\ee
The comoving time $t$ spans a finite range and the scale 
factor $R(t)\simeq |t-t_{i,f}|^{1\over 3}$ at early and late times.
Kaluza-Klein compactification of Einstein gravity in $D=5$ gives
the effective $D=4$ Lagrangian. It is a free scalar, minimally coupled 
to gravity:
\be
16\pi G_N{\cal L}= 
R^{(4)}_E -6g^{\mu\nu}\partial_\mu\sigma \partial_\nu\sigma
\ee
For a homogeneous field the ensuing energy momentum tensor describes 
matter with $\rho=p$. Friedmann's equations with this equation of state
of course reproduce the $|t-t_{i,f}|^{1\over 3}$ behavior found at 
early and late times.

Thus the $D=5$ vacuum black hole solution with a coordinate singularity 
-- the horizon -- and a physical singularity -- the origin --
becomes $D=4$ cosmology with scalar matter and physical 
(curvature) singularities at both initial and final times.
(For a related discussion of certain extremal black hole 
singularities see~\cite{gibbons94}. )  
{}From the perspective of the interior cosmology, either direction of 
temporal flow is equally good. The compactification of $y$ appears
innocuous for the interior solution, but its continuation 
beyond the horizon involves compactifying the real time $t$ (=$y$), 
so that there are closed time-like paths and normal causality 
conditions are violated in the exterior. Although these 
violations should not be accepted lightly, note that they are restricted 
to the necessarily exotic epoch `after' the end of time.  In any case,
since they arise in the natural continuation of an apparently
reasonable cosmology, to forbid them would involve a new cosmological veto.
At the horizon, $t$ is not a good coordinate and one must analyze using (say)
Kruskal coordinates $U,V$.  The periodic identification of $t$ becomes a
re-scaling $U \rightarrow \alpha U$, $V\rightarrow {1\over \alpha}V$.  
Though there is no curvature singularity, there is an unusual non-Hausdorff
topology at the fixed point $U=V=0$. 

A recurrent theme in recent work on duality is that the low energy 
field theory develops singularities when heavy states of the full 
theory become light \cite{seibergw,conifolds,scales}. In the 
re-interpretation of big crunch as a higher-dimensional horizon 
the relevant heavy states are the Kaluza-Klein modes from the 
compactified dimension: at late times they become light and important 
for the dynamics; ultimately, they take over the role of time. 

\bigskip 

This simplest model has several deficiencies, including a remaining
singularity at the origin of the $D=5$ black hole. In the following
we generalize the model and
\begin{itemize}
\item 
embed the cosmology in a solution that 
recognizably derives from string theory. 
\item
determine whether cosmologies derived from 
string theory naturally involve other kinds of matter.
\item
exhibit solutions in which the singularities are resolved
by a new scenario, using that  
D-branes are singular as classical 
solutions but non-singular in their physical behavior. 
Here the apparent singularity of the low energy field theory
arises because open string excitations, ignored in the effective
field theory, become light close to the D-branes.
\end{itemize}

Large classes of cosmological solutions, including some of those
considered below - but not their embedding inside black holes -
have been analyzed recently~\cite{lukas96a,kaloper,pope96b,lukas96b}.
Configurations with non-trivial RR-fields are of particular 
interest because they can be represented as simple 
conformal field theories~\cite{polch95a}. 
The starting point is the classical solution associated with the 
extremal Dirichlet p-brane~\cite{hs} (for a review of string solitons
see~\cite{duffreview}):
\bea
ds^2_S &=& e^{2\xi} ( -dy^2 + dx_1^2+ \cdots + dx_p^2 ) +
       e^{-2\xi}( dx_{p+1}^2 + \cdots + dx_9^2) 
\label{eqn:extmet} \\
e^{-2\phi_{10}} &=& e^{-2\xi (p-3)} 
\label{eqn:dilatonansatz}\\
F_{p+2} &=& \partial_\mu e^{4\xi} \: dy \wedge dx_1 \wedge \cdots
dx_p \wedge dx^\mu \\
e^{-4\xi} &=&  1 +  \frac{q}{r^2} ~,
\eea
where $F_{p+2}$ is the $p+2$ form RR-field strength.  We have written the
solution in the $D=10$ string frame, and anticipating compactification 
to $D=5$ have arranged a periodic array in $5$ small spatial dimensions.
Then the harmonic function $e^{-4\xi}$ assumes the form $r^{-2}$ 
where $r$ is a spatial 4-vector for any  $p$ ($\leq 5$). 
The extremal metric eq.~\ref{eqn:extmet}
has no horizon and therefore no interior; so we need its `black' 
generalization~\cite{pope96a,tseytlin96}. Inside the horizon it 
reads after renaming of variables
\bea
ds^2_S &=& e^{2\xi} ( e^{-4\sigma}~dy^2 + dx_1^2+ 
\cdots + dx_p^2 ) + \nonumber \\
      &+&   e^{-2\xi}  ( dx_{p+1}^2 + \cdots + dx_5^2)
      + e^{-2\xi} (-e^{4\sigma}d\tau^2+\tau^2 d\Omega_3^2) ~,
\label{eqn:blackdbrane}
\eea
where $e^{-4\xi}={q\over \tau^2}+1$ and as before
$e^{-4\sigma}= {\mu\over \tau^2}-1$, with $\mu$ a 
numerical `non-extremality' parameter. 
The field strength is also modified, as we will discuss presently.
Compactification -- including compactification 
of $y$ -- introduces an effective dilaton
$
e^{-2\phi_4}= e^{2\xi-2\sigma}
$
for all $p$. The $D=4$ Einstein metric again becomes 
eq.~\ref{eqn:basicline}. Thus the $D$-brane generates the 
same interior space-time as the Schwarzschild solution! 

To understand this result in more detail, consider
the metric eq.~\ref{eqn:blackdbrane} an {\it ansatz\/} with the
$D=4$ fields $\sigma$ and $\xi$ undetermined {\it a priori\/}.
Similarly, use eq.~\ref{eqn:dilatonansatz} as an {\it ansatz\/} for
the dilaton and introduce an independent scalar field $A$ that 
arises because the $(p+2)$-form field strength reduces to a 1-form, 
written as $\nabla A$. 
Then the $D=10$ string action
\be
16\pi G_N {\cal L}= e^{-2\phi_{10}} [ R^{(10)}_S+ 4 (\nabla\phi_{10})^2 ]
-{1\over 2(p+2)!}F^2_{p+2}~
\ee
gives rise to the effective $D=4$ Einstein frame action 
\be
16\pi G_N {\cal L}=  R^{(4)}_E- 6(\nabla\sigma)^2
-8(\nabla\xi)^2+8\nabla\xi\nabla\sigma
-{1\over 2}e^{4\sigma-8\xi}(\nabla A)^2 
\label{eqn:Daction}
\ee
after some calculation. The stress-energy tensor corresponding to 
this Lagrangian describes $\rho=p$ matter, just as was the case 
without $D$-branes. The equations of motion for the scalar fields 
can be written
\bea
\nabla^2\sigma &=&0 \label{eqn:sigmaeom}\\ 
\nabla_\mu (e^{4\sigma-8\xi}\nabla^\mu A )
\label{eqn:Aeom} &=&0 \\
4\nabla^2 \xi + e^{4\sigma-8\xi} (\nabla A)^2 &=&0~. 
\label{eqn:xieom}
\eea
We expect the solution 
\bea
ds_E^2 &=& - e^{2\sigma}d\tau^2 + e^{-2\sigma}\tau^2 d\Omega^2_3 
\label{eqn:dmetric} \\
e^{-4\sigma} &=&{\mu\over\tau^2}-1
\label{eqn:dsigma}\\
\partial_\tau A &=& {2Q\over \tau^3}e^{8\xi}~~;~Q^2=q(q+\mu)
\label{eqn:dA} \\
e^{-4\xi} &=&{q\over\tau^2}+1~.
\label{eqn:dxi} 
\eea
It is easy to verify that 
eqs.~\ref{eqn:dmetric}-\ref{eqn:dsigma} indeed satisfies 
eq.~\ref{eqn:sigmaeom} and that eq.~\ref{eqn:Aeom} 
requires $\partial_\tau A\propto\partial_\tau e^{4\xi}$ -- 
which is precisely the {\it ansatz} eq.~\ref{eqn:dA} above, 
with the proportionality constant, the charge, parametrized by $Q$. 
The final equation eq.~\ref{eqn:xieom} is non-trivial and is satisfied by 
virtue of cancellations that appear rather mysterious. 
Indeed in the energy--momentum  tensor
\be
16\pi G_N T_{\tau\tau}=6\dot{\sigma}^2
+8\dot{\xi}^2-8\dot{\xi}\dot{\sigma}+{1\over 2}e^{4\sigma-8\xi}{\dot A}^2
\ee
the first term is sufficient to satisfy the Friedmann equations
because $\xi=0$ is the solution without the D-brane. Therefore  
the subsequent terms must -- and do -- cancel. The higher dimensional 
physics leads us to define the scalar fields in such a way that only 
one of them couples to gravity, at least on-shell and for homogeneous
solutions. The additional fields are somewhat marginal from the 
$D=4$ perspective, in that they carry no net energy, but they 
parametrize inequivalent embeddings of the $D=4$ theory into a 
higher dimensional theory.



The $D=4$ effective action eq.~\ref{eqn:Daction} is independent of $p$
but the relation between the scalar fields $\xi$ and $\sigma$ 
and the higher dimensional geometry does depend on $p$.
It is straightforward to verify explicitly that branes with coupling 
to the NS fields are similarly described by the action 
eq.~\ref{eqn:Daction}, with yet another higher dimensional interpretation 
of the scalar fields. In fact, U-duality of the type II string theory
implies that {\it all\/} black branes must give the same $D=4$ 
Einstein metric. 

In all the models discussed so far, the big crunch (or big bang) 
singularity is resolved as a horizon in $D=5$ but the big 
bang (or big crunch) singularity remains
a curvature singularity.  In the microscopic theory for extremal
branes,
the source 
is a very large number 
of coincident D-branes. The non-extremality parameter $\mu$ can 
be introduced by considering non-supersymmetric excitations of this 
object~\cite{callan96a,horowitz96a}.  The physical nature of the 
apparent singularity can be explored using test D-brane (or closed
string) probes. Formally, the interactions
between the probe and the background are governed by
an annulus diagram.  It is finite for $\mu=0$ due to supersymmetry.
$\mu\neq 0$ can be modelled by including appropriate vertex operators
on the boundary of the annulus, and the diagram remains finite. 
Higher order corrections are also finite order by order, 
though not necessarily smaller.
The situation seems analogous to the case where supersymmetry
is broken by finite velocity~\cite{scales}.  
In that case, at large distances it is  
legitimate to truncate the implied sum over string states and 
describe the interaction as the field theoretic one expressed in 
the classical solution, while at 
small distances and very small velocities the interaction can be 
described in terms of exchange 
of a few open string states.  We expect that in our 
case $\mu \neq 0$, as for  velocities that are  not small, the interaction
remains
non-singular though at present it is not calculable.

\par

Regular black holes in $D=5$ are obtained by combining 
three black branes~\cite{horowitz96a,cvetic96a}. In fact, this 
solution is unique up to duality~\cite{hull96}. As a representative 
of this class, consider an NS string winding 
around the $1$ direction, a $D4$-brane wrapped around the $2345$ 
directions, and a $D0$ brane. The line element in $D=10$ string 
frame form of these three intersecting black branes is
\bea
ds^2_S &=& e^{4\xi_w+2\xi_0+2\xi_4-4\sigma}~dy^2 + 
e^{4\xi_w-2\xi_0-2\xi_4}dx_1^2 + \nonumber \\ 
 &+& e^{2\xi_4-2\xi_0} (dx^2_2+ dx^2_3+dx^2_4+dx^2_5)
      + e^{-2\xi_0-2\xi_4} (-e^{4\sigma}d\tau^2+\tau^2 d\Omega_3^2)
\nonumber \\
e^{-2\phi_{10}}&=&e^{6\xi_0-2\xi_4-4\xi_w} 
\label{eqn:dansatz}
\eea
where as before $e^{-4\sigma}$ and $e^{-4\xi_i}$ are harmonic 
functions. (Here $i=1,2,3$, by abuse of notation.)
Alternatively, consider this an {\it ansatz \/} and derive the
effective $D=4$ action 
\be
16\pi G_N {\cal L}_E^{(4)}= R^{(4)}_E- 6(\nabla\sigma)^2
+\sum_{i=1}^3 [8\nabla\sigma\nabla\xi_i-8(\nabla\xi_i)^2
- {1\over 2}e^{4\sigma-8\xi_i}(\nabla A_i)^2]
\label{eqn:Dact}
\ee
It is a consequence of U-duality that the fields $\xi_i$ appear
symmetrically despite the {\it ansatz\/} eq.~\ref{eqn:dansatz}.
The equations of motion 
factorize appropriately so that their solution indeed becomes the 
expected one, {\it i.e.\/}
eqs.~\ref{eqn:dmetric}--\ref{eqn:dxi} (with three identical copies
of eqs.~\ref{eqn:dA}--\ref{eqn:dxi} ) .

To make contact with more familiar ideas, let us consider the special
case where the 3 charges are equal.
Then upon reducing to 5 dimensions one finds
\be
ds^2 = e^{8\xi-4\sigma}~dy^2 
  + e^{-4\xi} (-e^{4\sigma}d\tau^2+\tau^2 d\Omega_3^2)
\ee
and a time independent dilaton; so string frame agrees with Einstein frame. 
Introducing black hole notation the line element becomes
\be
ds^2 = -{1\over (1+{q\over r^2})^2} (1-{\mu\over r^2})dt^2
+(1+{q\over r^2}){1\over 1-{\mu\over r^2} }dr^2 +(1+{q\over r^2})r^2 
d\Omega^2_3~,
\ee
and finally in terms of the geometrical radius $R^2=r^2+q$
(which governs the volume of the overlying spheres) 
\be
ds^2 = -(1-{R^2_{+}\over R^2})(1-{R^2_{-}\over R^2})dt^2 +
{1\over (1-{R^2_{+}\over R^2})(1-{R^2_{-}\over R^2})}dR^2 +
R^2 d\Omega^2_3~,
\ee
where $R^2_{+}=q+\mu$ and $R^2_{-}=q$ specify the outer and inner
horizons respectively. Also $2G_N M=2q+\mu$ and 
$R^2_{+}R^2_{-}=Q^2$.  In this form it is clear that
the horizons are
indeed coordinate singularities and that the origin $R=0$ is a curvature
singularity, even in the extremal case. 

Returning to the general case the $D=4$ cosmological geometry is 
again the familiar one, which displays remarkable robustness. 
The dilaton behaves as
$e^{-2\phi_4}=e^{2\xi_0+2\xi_4-2\sigma}$ and approaches zero at the inner 
horizon/origin of the universe. This suggests that the coupling 
constant for string perturbation theory diverges. It appears now, 
however, to be a pseudo-singularity 
caused by an illegal truncation of the theory: specifically, 
the volume of the universe goes to zero because the length of the 5th 
dimension goes to zero. In $D=5$ all curvatures remain bounded and
indeed small near the horizon  
provided  $\mu \gg l^2_P$ and   $q_i \gg l_P^2$ where $l_P$
is the $D=5$ Planck length.  The first condition is automatic for
a macroscopic universe.
The second requires that the inner horizon is well separated from 
the singularities.  It will  be satisfied when all the $Q_i$ are large.  
If it is not satisfied, the singularity is
not resolved within the field theory approach.  However, it is
plausible that the space-time singularity, which acts as the source of
mass and charge, marks the approximate space-time locus (to the extent this
notion makes sense) of the region where the brane intersection
resides, and where an adequate description of the physics must take into
account the open-string degrees of freedom which become light there.

\par 
The proposed cosmology is one where, at early times, the evolution is
dominated by the Lagrangian eq. \ref{eqn:Dact}.  Because of their stiff
equation of state, the effective scalars will tend to dominate other forms
of matter close to the apparent singularities.
One or more of these scalars diverge close to the origin of time; so
the $D=4$ effective field theory
breaks down and must be replaced by something more exotic. Depending
on the details (the number of non-vanishing $\xi_i$) the natural
description becomes $D=5$ effective field theory with closed time-like 
loops, or an open string effective field theory without a conventional 
interpretation in terms of geometric quantities. In both cases the
transition is perfectly smooth. 

\par 

Our models are quite schematic.  Even
if one of them 
were to give a valid first indication for physics very near the
apparent big bang singularity, much
more complex dynamics will come into play afterwards.
As a simple example of the possibilities, note that if 
the crossover at late times is before
the integration constants $\pm 1$ in the harmonic functions become
significant in the metric the universe could still be flat, or even open. 
The relevant solutions generalize those found in~\cite{behrndt}:
\bea
ds_E^2 &=& - e^{2\sigma}d\tau^2 + e^{-2\sigma}\tau^2 d\Omega^2_3 \\
e^{-4\sigma} &=&{\mu\over\tau^2}-k\\
e^{-4\xi_i} &=&{q_i\over\tau^2}+1 \\
\partial_\tau A_i &=& {2Q_i\over \tau^3}e^{8\xi}~~;~Q_i^2=q_i(kq_i+\mu)
\eea
Here $k=-1,0,1$. The solutions with $k=-1,0$ can not be interpreted 
as black holes.  

{\bf Acknowledgments:} 
We would like to thank A. Lukas and B. Ovrut for discussions.  
F.L. is supported in part by DOE grant DE-AC02-76-ERO-3071. 
F.W. is supported in part by DOE grant DE-FG02-90ER-40542


\end{document}